\documentclass[twocolumn]{aastex61}

\usepackage{url}

\newcommand\gx{GX 339$-$4}

\newcommand\msun {$M_{\odot}$}

\newcommand\flxd{erg cm$^{-2}$ s$^{-1}$ \AA$^{-1}$}

\newcommand\vel{km s$^{-1}$}

\newcommand\h{{\it H}-band}

\newcommand\jband{{\it J}-band}

\def\porb{$1.7587 \pm 0.0005$ d}
\def\gam{$26 \pm 2$ km s$^{-1}$}
\def\k2{$219 \pm 3$ km s$^{-1}$}
\def\vsini{$64 \pm 8$ km s$^{-1}$}
\def\qmeas{$0.18 \pm 0.05$}
\def\fm{$1.91 \pm 0.08~M_{\odot}$}
\def\bhmax{$9.5~M_{\odot}$}
\def\bhmin{$2.3~M_{\odot}$}
\def\mininc{$37^\circ$}
\def\maxinc{$78^\circ$}

\received{}
\revised{}
\accepted{Aug 9 2017}
\submitjournal{ApJ}

\shorttitle{Mass function of \gx}
\shortauthors{Heida et al.}

\begin{document}

\title{The mass function of \gx{} from spectroscopic observations of its donor star\footnote{Based on ESO programme IDs 097.D-0915 and 297.D-5048}}

\correspondingauthor{M.~Heida}
\email{mheida@caltech.edu}

\author[0000-0002-1082-7496]{M.~Heida}
\affil{Cahill Center for Astronomy and Astrophysics, California Institute of Technology \\
1200 California Blv., Pasadena, CA 91125, USA}

\author[0000-0001-5679-0695]{P.G.~Jonker}
\affil{SRON Netherlands Institute for Space Research\\
Sorbonnelaan 2, 3584 CA Utrecht, The Netherlands}
\affil{Department of Astrophysics/IMAPP, Radboud University\\
P.O.~Box 9010, 6500 GL Nijmegen, The Netherlands}

\author{M.A.P.~Torres}
\affil{Instituto de Astrof\'{i}sica de Canarias\\ 
E-38205 La Laguna, S/C de Tenerife, Spain}
\affil{Departamento de Astrof\'{i}sica, Universidad de La Laguna\\ 
E-38206 La Laguna, S/C de Tenerife, Spain}
\affil{SRON Netherlands Institute for Space Research\\
Sorbonnelaan 2, 3584 CA Utrecht, The Netherlands}
\affil{Department of Astrophysics/IMAPP, Radboud University\\
P.O.~Box 9010, 6500 GL Nijmegen, The Netherlands}

\author{A.~Chiavassa}
\affil{Universit\'e C\^ote d'Azur, Observatoire de la C\^ote d'Azur, CNRS\\
Lagrange, CS 34229, 06304 Nice Cedex 4, France}



\begin{abstract}
We obtained 16 VLT/X-shooter observations of \gx{} in quiescence in the period May -- September 2016 and detected absorption lines from the donor star in its NIR spectrum. This allows us to measure the radial velocity curve and projected rotational velocity of the donor for the first time. We confirm the 1.76 day orbital period and we find that $K_2$ = \k2, $\gamma$ = \gam{} and $v \sin i$ = \vsini. From these values we compute a mass function $f(M)$ = \fm, a factor $\sim 3$ lower than previously reported, and a mass ratio $q$ = \qmeas. We confirm the donor is a K-type star and estimate that it contributes $\sim 45-50\%$ of the light in the $J$- and \h. 
We constrain the binary inclination to \mininc $< i <$ \maxinc{} and the black hole mass to \bhmin $< M_\mathrm{BH} <$ \bhmax. \gx{} may therefore be the first black hole to fall in the `mass-gap' of $2-5$ \msun.  
\end{abstract}


\section{Introduction} 
\gx{} is a low mass X-ray binary (LMXB) system that has been studied extensively since its discovery \citep[e.g.][]{markert73, miyamoto91, mendez97, gallo04, plant15}. It shows relatively frequent outbursts (four in the last decade); all four X-ray states typically seen in X-ray binaries (XRBs) have been detected in this system, and X-ray observations of this system have been very important in shaping the theory of LMXBs \citep[e.g.][]{miyamoto95, homan05, plant14}. The relation between X-ray and radio emission from XRBs also relies heavily on observations of \gx{} \citep[e.g.][]{corbel13, gallo14}. 

However, the mass of the compact object in \gx{} is not known. Although it has long been suspected to harbor a black hole (BH) \citep{cowley87,mcclintock06}, dynamical evidence for this was long missing because its optical spectrum in quiescence is dominated by the accretion disc \citep{shahbaz01}. %, because the system never really goes into quiescence \citep[e.g.][]{shahbaz01}. 
\citet{hynes03} were the first to detect dynamical evidence for a BH from CIII/NIII Bowen emission lines modulated on an orbital period of 1.7557 days. These lines are suspected to originate in the irradiated atmosphere of the donor star. They found a lower limit on the radial velocity semi-amplitude ($K_2$) of $317 \pm 10$ \vel, corresponding to a mass function $f(M) = 5.8 \pm 0.5$ \msun. Their reported orbital period was later detected in X-rays as well \citep{levine06}. \citet{munoz-darias08} refined the \citet{hynes03} result by applying the K-correction and suggested that the donor star is a stripped giant with an effective temperature of $\sim 4500-5000$K. 

The mass function reported by \citet{hynes03} was surprisingly large given that the inclination of the system is believed to be low based on the small separation of the double peaked emission lines (cf.~\citealt{wu01}). To obtain a reliable mass function for \gx, it is necessary to measure the radial velocity (RV) curve using absorption lines from the photosphere of the donor star. As these have not been detected in its visible spectrum because of the bright accretion disc, we used the very wide spectral range of X-shooter \citep{vernet11} on the Very Large Telescope (VLT) to search for absorption lines in the near-infrared (NIR) while \gx{} was in a quiescent phase in the summer of 2016. A similar approach led to the measurement of the mass function for GRS 1915+105, a system that is too obscured to be observed at visible wavelengths \citep[although for GRS 1915, the RV curve was measured while the system was in outburst;][]{greiner01,steeghs13}.
Here we present our measurement of the RV curve of \gx{} using stellar absorption lines detected in the $J$- and \h. Section \ref{par:obs} describes our observations and data reduction. In Section \ref{par:res} we describe our RV measurements and our analysis of the average spectrum to determine the rotational velocity ($v\sin i$) of the donor star. We discuss the implications for the mass function, our estimate of the mass ratio, and the distance to \gx{} and present our conclusions in Section \ref{par:conc}. All errors quoted in this paper are $1-\sigma$ single parameter errors, and limits are 95\% confidence limits. All errors on derived quantities are calculated using Monte Carlo error analysis. Quoted wavelengths are in air.

\section{Observations and data reduction}\label{par:obs}
\subsection{X-shooter spectra of \gx}
We obtained 16 VLT/X-shooter spectra of \gx{} in service mode while the system was in quiescence: 1 pilot observation on May 22, 2016 (run code 097.D-0915) and subsequently 15 observations between August 10 and September 8, 2016 granted as director's discretionary time (run code 297.D-5048; see Table \ref{tab:obs} for a log of the observations). Every spectrum was obtained in one observing block (OB) including the target acquisition and one ABBA nod pattern, using the XSHOOTER\_slt\_obs\_AutoNodOnSlit template and a nod throw of 5''. Individual exposure times are 250, 200 and 275 s for the UVB, VIS and NIR arms, respectively. This yields total integration times of 1000, 800 and 1100 s for the respective arms for each of the 16 observations. We use the 1.0'' slit in the UVB arm, the 0.9'' slit in the VIS arm and the 0.9JH'' slit (with K-band blocking filter) in the NIR arm. This setup was chosen to maximize signal to noise (S/N) in the $J$- and $H$-bands. The slitwidth-limited resolution R$ = \lambda/\Delta \lambda$ is 4350 in the UVB arm, 7410 in the VIS arm and 5410 in the NIR arm.
In some observing nights, more than one OB was executed. In these cases the target was reacquired at the beginning of each OB. A telluric standard star was observed for every science spectrum. Sky conditions were photometric on 2016-08-14 and clear on all other nights. 

\begin{table*}
  \centering
 \caption{Log of our VLT/X-shooter observations of \gx. }\label{tab:obs}
\begin{tabular}{cccccc}
 \hline
 \hline
 UT date & MJD mid time & Sky\footnote{The sky conditions in all observations were either clear (CLR) or photometric (PHO).} & VIS IQ\footnote{The image quality (IQ) is measured as the FWHM of the spatial profile of the 2D spectrum in the $V$, $R$, $J$, and $H$-bands.} & NIR IQ & RV\\
 (yyyy-mm-dd) &  &  & ($V$ - $R$, arcsec) & ($J$ - $H$, arcsec) & (\vel) \\
  \hline
2016-05-22 & 57530.7647627 & CLR & 0.75-0.8 & 0.8-0.7 & $-190 \pm 30$ \\
2016-08-09 & 57610.4930110 & CLR & 1.0 & 1.0-0.85 & $183 \pm 11$ \\
2016-08-10 & 57611.4814135 & CLR & 0.9 & 0.75-0.65 & $-161 \pm 15$ \\
2016-08-12 & 57613.4822289 & CLR & 0.9-0.8 & 0.8-0.7 & $-164 \pm 8$ \\
2016-08-14 & 57614.5926091 & PHO & 1.0 & 1.0-0.85 & $106 \pm 5$ \\
 & 57614.6310668 & PHO & 0.85-0.95 & 0.9-0.8 & $76 \pm 5$ \\
2016-08-30 & 57631.4852136 & CLR & 0.85-0.95 & 0.8-0.7 & $87 \pm 5$ \\
2016-08-31 & 57631.5239705 & CLR & 0.9 & 0.8-0.7 & $131 \pm 12$ \\
 & 57631.5558336 & CLR & 0.85-1.0 & 0.95-0.8 & $125 \pm 15$ \\
2016-09-03 & 57634.5215575 & CLR & 0.9 & 0.9-0.8 & $-175 \pm 9$ \\
 & 57634.5612928 & CLR & 0.8 & 0.8-0.65 & $-197 \pm 5$ \\
2016-09-04 & 57636.4847256 & CLR & 0.9-0.8 & 0.75-0.65 & $-102 \pm 9$ \\
2016-09-05 & 57637.4787554 & CLR & 0.8 & 0.7-0.65 & $81 \pm 13$ \\
2016-09-06 & 57637.5182959 & CLR & 0.9 & 0.75-0.7 & $47 \pm 5$ \\
 & 57637.5527546 & CLR & 1.0 & 0.85-0.8 & $37 \pm 9$ \\
2016-09-07 & 57639.4869854 & CLR & 0.7-0.8 & 0.65-0.6 & $-89 \pm 8$ \\ 
  \hline  
 \end{tabular}
\end{table*}

For all observations we use a sky position angle of 119$^\circ$ so that both the counterpart to \gx{} and a nearby star (star B in \citealt{shahbaz01}) fall in the slits of the three arms. The separation between \gx{} and star B is only $1.07''$, so we require a seeing of $< 0.9''$ (as measured in the V-band) to separate the two spectra. This condition was met in most observations. Because in the NIR arm the seeing is smaller than the slit width in all observations, the resolution is set by the seeing rather than by the slit. This means that the effective resolution varies between the spectra and is somewhat higher than the slit-limited resolution of R = 5410, or $\sim 56$ \vel{} at 1.65 $\mu$m. To quantify this, we measure the full width at half maximum (FWHM) of the spectra by fitting two Gaussians to the spatial profile of star B and \gx{} at several wavelengths (see Table \ref{tab:obs}). The spectra taken under the best seeing conditions have a FWHM in the \h{} of $\sim 0.65''$, which corresponds to a resolution of R = 7490 or $\sim 40$ \vel.

As expected, there is hardly any signal in the UVB arm. The S/N in the VIS arm is rather low ($\lesssim 5$) but the H$\alpha$ line is clearly detected. The NIR arm has the highest continuum S/N ratio in all observations, varying from $\sim 5-15$ depending on the seeing. We analyze the data from the VIS and NIR arms only.

We process the data of \gx{} and the telluric standards using the X-shooter pipeline in the {\sc Reflex} environment \citep{freudling13}. This produces a flat-fielded, sky-subtracted, rectified, wavelength- and flux-calibrated 2D image and a 1D extracted spectrum. We also use the {\sc Starlink} package {\sc Figaro} to optimally extract the \gx{} spectra with the tasks {\it profile} and {\it optextract} \citep{horne86}, using a non-symmetric extraction region to avoid as much as possible contamination from star B. Comparing the spectra extracted in these different ways, we find that the optimal extraction significantly reduces the errors for the NIR arm spectra, while it does not improve the VIS arm spectra that are dominated by emission lines. Therefore we use the optimally extracted spectra in the NIR arm and the spectra provided by the pipeline for the VIS arm. For the telluric standards we take the normally extracted spectra provided by the pipeline.

The size of the extraction aperture for the NIR spectra is determined by the proximity of star B, not by the width of the spatial profile. This causes a loss of NIR flux that is most significant in the spectra taken under the worst seeing conditions. In combination with the fact that most spectra were taken under clear, not photometric conditions this means that the absolute flux calibration of our spectra is not accurate.

For the telluric correction of the NIR arm data, we use the telluric calibration package {\sc Molecfit} \citep{kausch15,smette15}. First we fit an atmospheric model to the telluric standard stars, to determine the concentrations of H$_2$O, CO$_2$, CH$_4$ and O$_2$. We then fit an atmospheric model to the \gx{} spectra, keeping the concentrations fixed, and apply the telluric correction.

\subsection{Templates}
To measure the RVs and rotational broadening of the \gx{} spectra we compare them with template spectra. We use both spectra of Galactic stars observed with X-shooter and synthetic spectra from model atmospheres at different temperatures as templates.

We retrieved the raw data of several K-stars and one late G-type star from the ESO archive (see Table \ref{tab:xsl}). These stars were observed with X-shooter as part of the X-shooter Spectral Library (XSL) project, with the $0.9''$ NIR arm slit as for our \gx{} data but without the K-band blocking filter \citep[][program ID 085.B-0751]{chen14}. All were observed with seeing $> 0.9''$ in the NIR so their spectral resolution is determined by the slit width. The XSL contains only one K-type subgiant; to obtain templates spanning a range of spectral types we also include several giant stars. We reduce the NIR arm data in the same way as for the telluric standard stars, using the X-shooter pipeline in {\sc Reflex} to extract the spectra and {\sc Molecfit} to correct for telluric absorption. 

\begin{table}
  \centering
 \caption{List of Galactic stars used as templates}\label{tab:xsl}
\begin{tabular}{lc}
 \hline
 \hline
 Star & Spectral type  \\
   \hline
HD 83212 &  G8III \\
HD 170820 &  K0III \\
HD 165438 & K1IV \\
HD 37763 & K2III \\
HD 175545 & K2III \\
HD 65354 & K3III \\
 \hline  
 \end{tabular}
\end{table}

In addition we use three model atmospheres with $\log{g} = 2.5$, solar metallicity, and effective temperatures of $3938$, $4500$ and $5000$ K. These were produced using the state-of-the-art realistic three-dimensional radiative hydrodynamical (RHD) simulations of stellar convection carried out with the Stagger-code \citep{nordlund09}. These simulations cover a large part of the HR diagram including the evolutionary phases from the main sequence over the turnoff up to the red-giant branch for low-mass stars \citep{magic13}. These simulations have been used to compute synthetic spectra in the range of X-shooter with the multidimensional pure-LTE radiative transfer code Optim3D \citep{chiavassa09}. The code takes into account the Doppler shifts occurring due to convective motions and solves monochromatically for the emerging intensity including extensive atomic and molecular continuum and line opacity data from UV to far-IR. We smooth the high-resolution RHD simulations both to the slitwidth-limited X-shooter resolution ($56$ \vel) and to the highest resolution present among our \gx{} observations ($40$ \vel) and rebin the two sets of synthetic spectra to the same velocity scale as the data.

\section{Analysis and results}\label{par:res}
\subsection{The RV curve of \gx}\label{par:rv}
\begin{figure*}
\includegraphics[width=\textwidth]{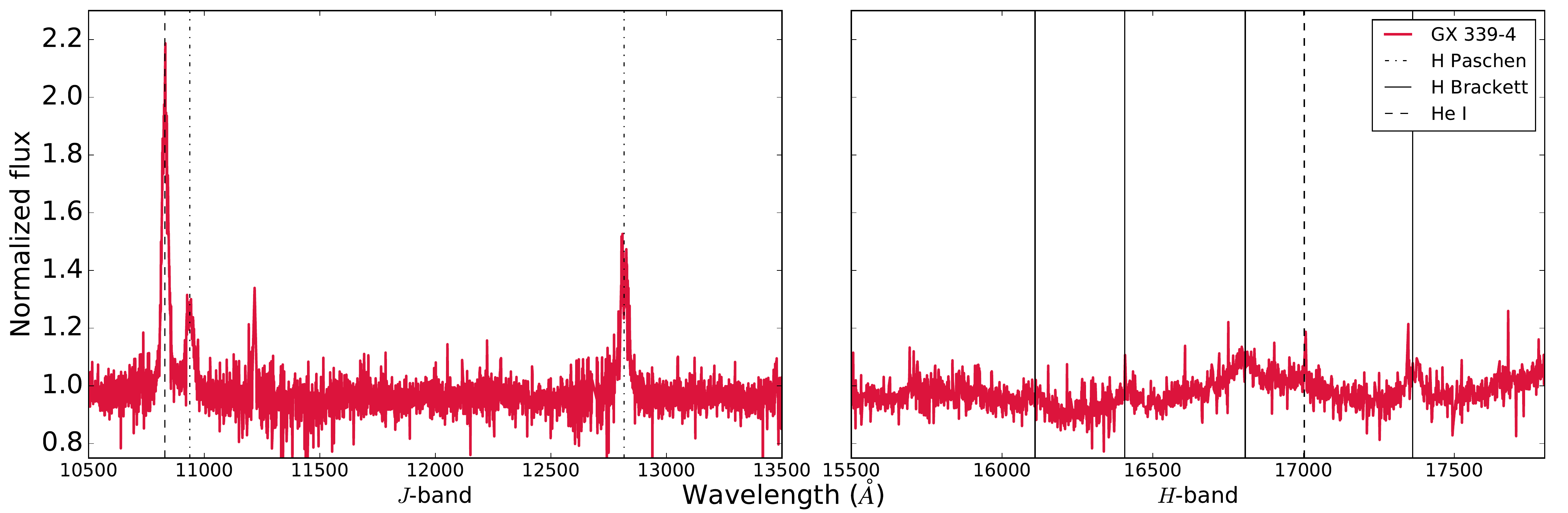}
\caption{The average, normalized $J$- and \h{} spectrum of \gx. Here the spectra are averaged without removing radial velocities, to emphasize the emission lines. The \jband{} has the strongest emission lines (of He I and H Paschen), while the \h{} is dominated by weaker, broad Brackett emission lines.}\label{fig:nirem}
\end{figure*}

We use Tom Marsh' program {\sc Molly}\footnote{\url{http://www2.warwick.ac.uk/fac/sci/physics/research/astro/people/marsh/software/}} to further analyze the data. First we apply {\it hfix} to convert to heliocentric velocities and we mask noisy pixels (mainly due to residuals from strong telluric emission lines). 
For the NIR data, we use {\it vbin} to rebin all spectra to a common velocity spacing of $15$ \vel{} per pixel, the average velocity dispersion of X-shooter in the $J$- and \h. Absorption lines corresponding to Mg {\sc i} ($1.71\mu$m) and Al {\sc i} (at $1.31\mu$m) are visible in the individual spectra, indicating that the donor star is a K-type star \citep[e.g.][see Section \ref{par:spec}]{meyer98,wallace00}.
To prepare the spectra for cross-correlation, we first scale them by dividing through a constant and then subtract a spline-fit. 
K-star spectra have many absorption lines which makes defining the continuum level difficult. We determine the normalization factor by fitting a constant to a relatively clean region of the spectrum close to the strong Mg {\sc i} line at 1.71$\mu$m. For the spline fit, we mask the regions in the \jband{} and between the $J$- and \h{} that are severely affected by telluric absorption, as well as the H and He emission lines that are present in the \gx{} spectra (see Figure \ref{fig:nirem}).

We prepare the templates in the same way as the \gx{} spectra. In addition we cross-correlate the spectra of the templates stars observed with X-shooter with the RHD simulations to obtain accurate radial velocities, and shift the template spectra to remove their intrinsic radial velocity.
We then cross-correlate the 16 individual \gx{} spectra with the different templates using the task {\it xcor}, excluding regions with high noise levels and emission lines. All three RHD simulations, at both resolutions, and the XSL stars yield consistent RVs. The highest values for the cross-correlation function are found for the $T = 3938$ K model smoothed to $56.6$ \vel, hence the results in this paper are from the cross-correlation with that template. For five \gx{} spectra with lower S/N we find that only considering the regions around the strongest absorption lines in the cross-correlation significantly improves the signal. The highest S/N spectra, and low S/N spectra with many masked pixels in these regions, do not benefit from this approach --- for those we use the RV found by cross-correlating the full selected spectral range. 

{\it xcor} gives only a purely statistical error that underestimates the real uncertainty in the RV measurement. To better constrain the uncertainty we use the {\it boot} command in {\sc Molly} to produce 1000 bootstrapped copies of every spectrum which we then cross-correlate with the model spectrum. We fit a Gaussian to the resulting distribution of RVs, and adopt the mean and standard deviation of that Gaussian as reliable estimates of the RV and uncertainty. The results are listed in Table \ref{tab:obs}; the $1-\sigma$ error includes the uncertainty in the wavelength calibration of X-shooter (2 \vel).

\begin{figure}
\includegraphics[width=0.5\textwidth]{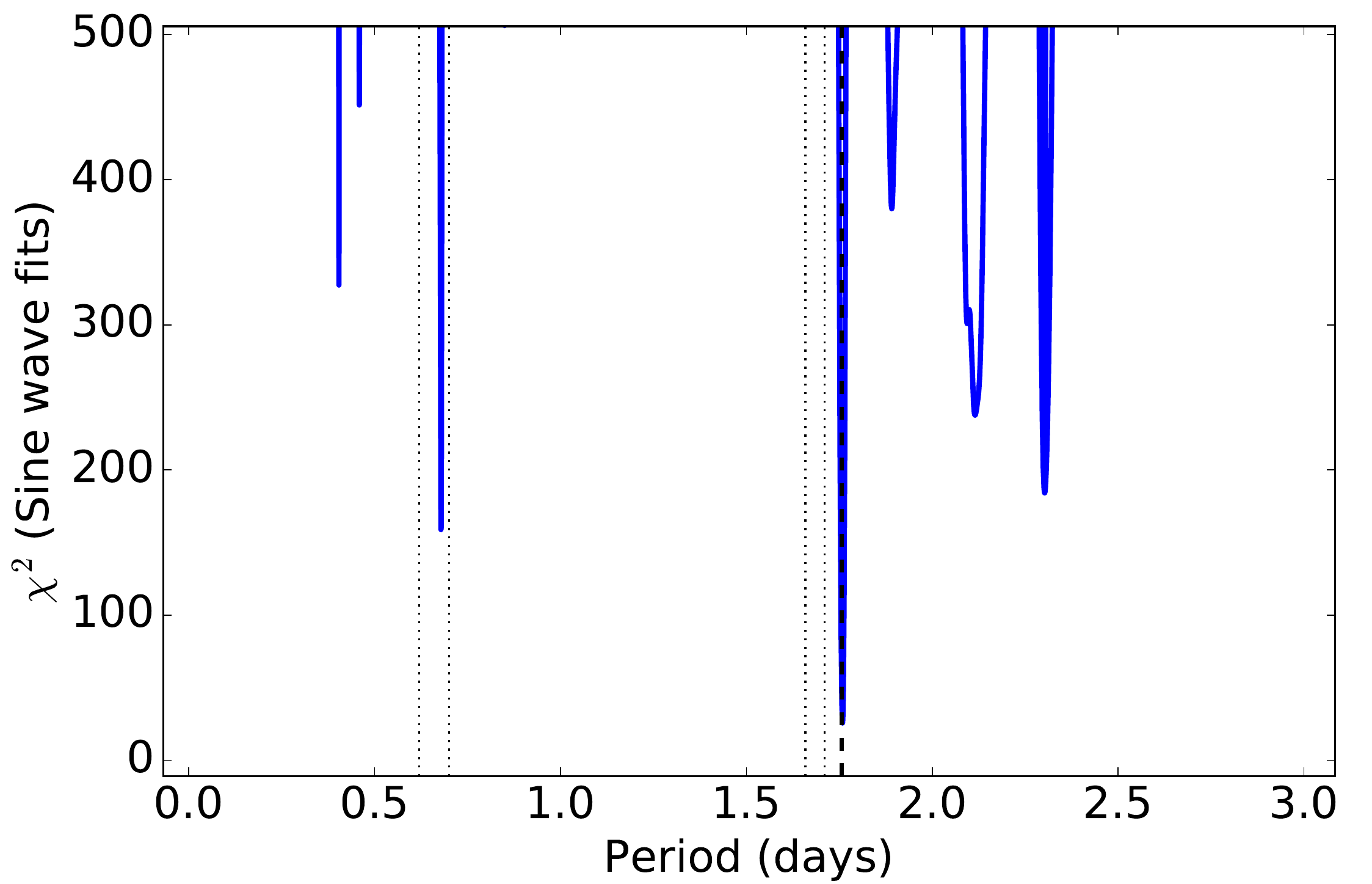}
\caption{$\chi^2$ of fits of a sinusoid to the RV points, as a function of period. The dashed line indicates the 1.7557 d period that was preferred by \citet{hynes03}; the dotted lines indicate different periods of 0.62 \citep{callanan92}, 0.7 \citep{cowley02}, 1.6584 and 1.714 d \citep{hynes03} that have been suggested as the orbital period. We find a best-fitting period of \porb, with a minimum $\chi^2$ of 25.6 for 13 degrees of freedom.}\label{fig:porbs}
\end{figure}

\begin{figure}
\includegraphics[width=0.5\textwidth]{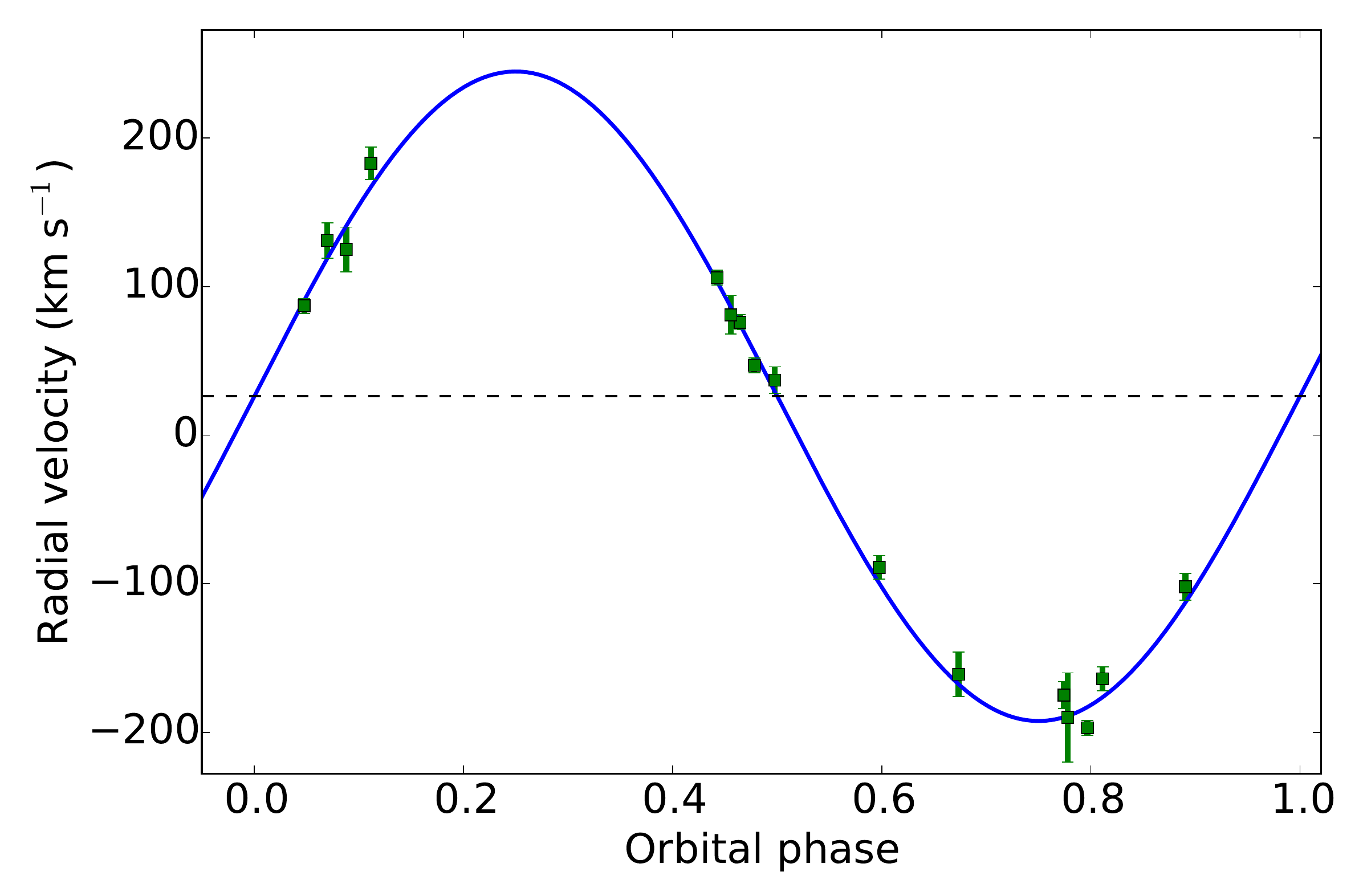}
\caption{Our 16 RV points folded on the preferred period of 1.7587 d. The best-fitting sinusoid has an offset $\gamma$ = \gam{} (indicated by the dashed line) and a semi-amplitude $K_2$ = \k2. Phase 0 occurs at MJD $57529.397 \pm 0.003$.}\label{fig:rvcurve}
\end{figure}

To determine the orbital period we fit sinusoids of the form $\gamma + K_2 \sin(2\pi (t - T_0)/P)$ for a range of periods and calculate $\chi^2$ (see Figure \ref{fig:porbs}). We find that the minimum $\chi^2$ is 25.6 (with 13 degrees of freedom) for a period of $P =$ \porb, where the $1-\sigma$ uncertainty is set by the change in $P$ for which $\Delta \chi^2 = 1$. Our spectroscopic orbital period confirms the orbital period reported by \citet{hynes03} and \citet{levine06}. 
The resulting RV curve is plotted in Figure \ref{fig:rvcurve}. 
The best-fit parameters are $\gamma$ = \gam, measured with respect to the heliocentric velocity, $K_2$ = \k2, and the time of inferior conjunction of the donor star $T_0 =$ MJD $57529.397 \pm 0.003$.

\begin{figure}
\includegraphics[width=0.5\textwidth]{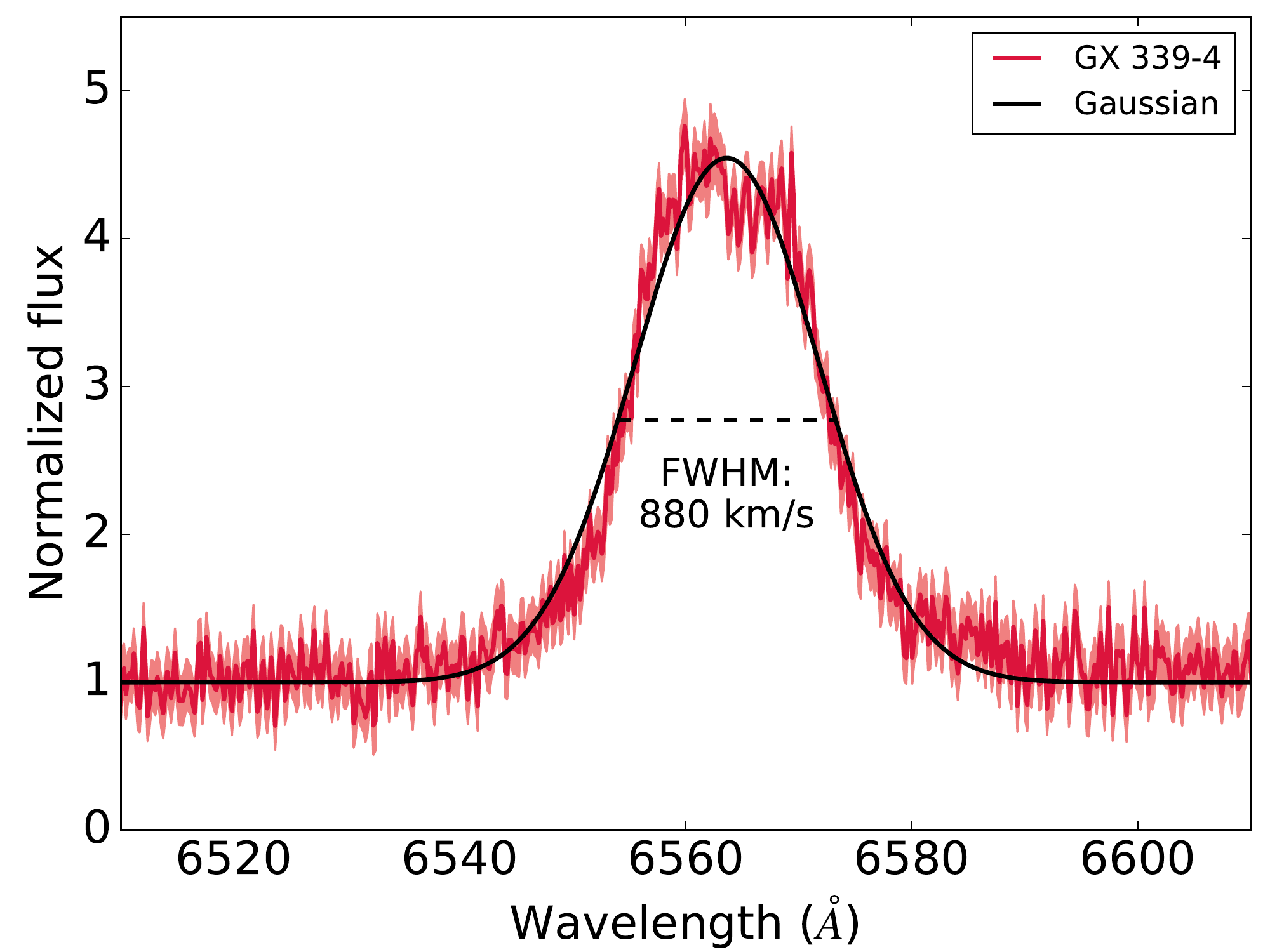}
\caption{The H$\alpha$ emission line in our average normalized spectrum of \gx{} (red line) and the best-fit Gaussian (black line). Note that although we show the average profile, the FWHM was obtained by fitting a Gaussian to the H$\alpha$ line in the individual spectra.}\label{fig:halpha}
\end{figure}

In addition we evaluate $K_2$ using the relation between this orbital parameter and the FWHM of the H$\alpha$ line \citep{casares15}. The H$\alpha$ line in our \gx{} spectra is double-peaked. Following the approach of \citet{casares15}, we fit a single Gaussian profile to the H$\alpha$ emission line in each VIS spectrum with {\it mgfit}. We find an average FWHM of $880 \pm 20$ \vel (corrected for the instrumental resolution; see Figure \ref{fig:halpha}). This corresponds to $K_2 = 206 \pm 12$ \vel{} according to the \citet{casares15} relation.

\subsection{Projected rotational velocity}\label{par:spec}
\begin{figure*}
\includegraphics[width=\textwidth]{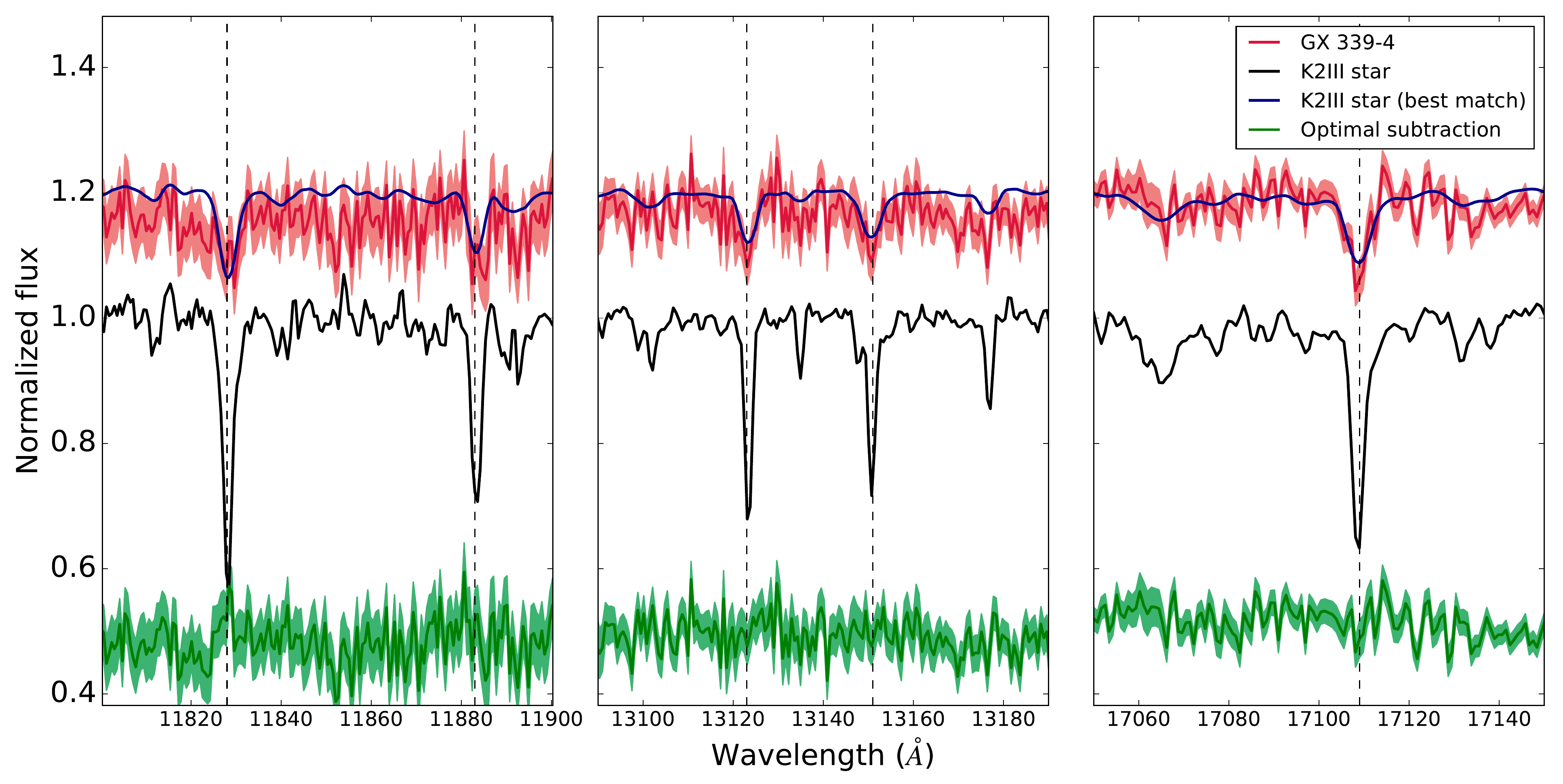}
\caption{Normalized average spectra of \gx{} (red line), the K2III star HD 175545 (black line), and HD 175545 broadened by 64 \vel{} and including a $50\%$ contribution from an accretion disc (blue line). The spectra of \gx{} and  the best-matched template are shifted upwards by 0.2 for clarity. Shown are the spectral regions around the Mg {\sc i} line at 1.1828 $\mu$m and the Fe {\sc i} line at 1.1883 $\mu$m, the Al {\sc i} doublet at 1.3123, 1.3151 $\mu$m, and the Mg {\sc i} line at 1.7109 $\mu$m; the lines are indicated by dashed lines. 
Also plotted are the residuals after optimal subtraction of the best-matched template from the \gx{} spectrum (green line).}\label{fig:spec}
\end{figure*}

To obtain a higher S/N spectrum for determining both the projected rotational velocity and spectral type of the donor star, we normalize the individual NIR spectra by dividing them through a spline fit, correct for the RVs of the individual spectra and average them, weighted by their mean (S/N)$^2$. In the resulting spectrum several absorption lines are detected, mainly due to Al {\sc i} and Mg {\sc i} (see Figure \ref{fig:spec}). The lines are weaker than those in the model spectra, indicating that the accretion disc still contributes significantly to the detected NIR continuum flux, and broadened because of the rotation of the tidally locked star. 

To determine the projected rotational velocity $v\sin i$ of the donor star, and the contribution of the accretion disc to the NIR emission, we use the optimal subtraction method \citep{marsh94}. In {\sc Molly}, we broaden all templates by a range of velocities (0 -- 100 \vel{} in steps of 1 \vel) with {\it rbroad} and then use {\it optsub} with a linear limb darkening coefficient of 0.75 to subtract the broadened models from the average \gx{} spectrum. For the optimal subtraction we use small spectral regions containing the five strongest absorption lines in the \gx{} spectrum. {\it optsub} subtracts a constant times the template from the \gx{} spectrum, adjusting the constant to minimise the residual scatter between the spectra. The scatter is measured by computing the $\chi^2$ between the residual spectrum and a smoothed version of itself, produced by convolving the subtracted spectrum with a Gaussian with a FWHM of 25 pixels. 

\begin{figure}
\includegraphics[width=0.5\textwidth]{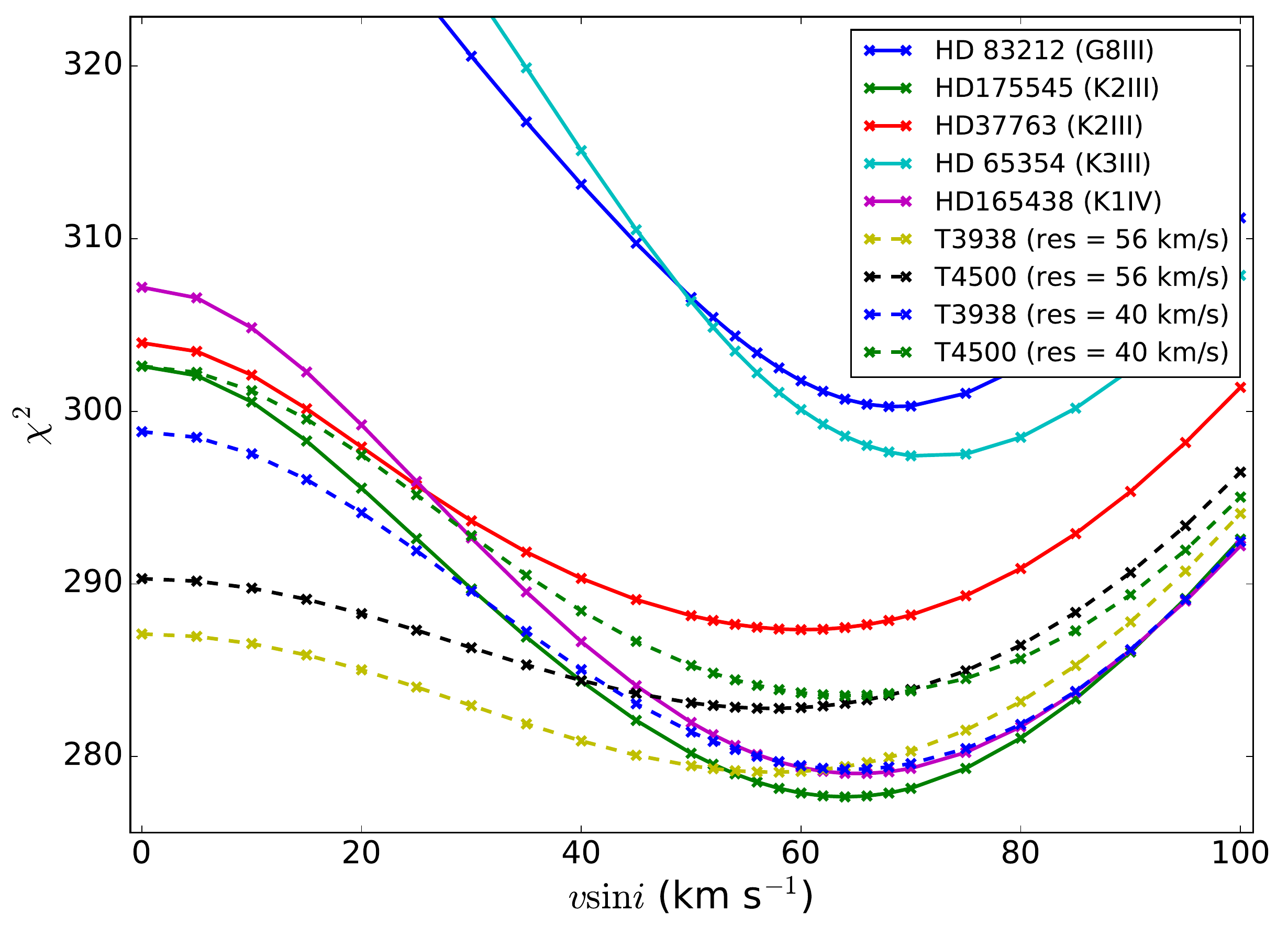}
\caption{$\chi^2$ distribution as a function of $v \sin i$ (361 degrees of freedom) obtained by optimal subtraction of templates from the average \gx{} spectrum. Results are shown for optimal subtraction of the $T=3938$ K and $T=4500$ K RHD simulations at resolutions of both 40 and 56 \vel, as well as for several stars observed with X-shooter for the XSL project. The best match is found for the K1IV and K2III stars and the $T=3938$ K RHD simulation.}\label{fig:chis}
\end{figure}

Apart from the two worst-matched templates, all $\chi^2$ distributions yield consistent $v \sin i$ values, with only small differences between the results found by using the higher and lower resolution RHD simulations (see Figure \ref{fig:chis}). This confirms that we are resolving the stellar absorption lines with our instrumental set-up under the prevailing seeing conditions. We find that using a different limb darkening factor of 0.5 influences the $v \sin i$ measurement by $< 5$ \vel. The best-fitting templates are the $T=3938$ K RHD simulation, a K2III star and the K1IV star; the K2III star HD 175545 has the lowest minimal $\chi^2$ of 278 (with 361 degrees of freedom) for an optimum factor of $0.47 \pm 0.04$. This template, and the residuals of the optimal subtraction, are shown in Figure \ref{fig:spec}. To obtain a reliable estimate of the uncertainty we follow the bootstrapping approach by \citet{steeghs07}: we produce 1000 bootstrapped copies of the average \gx{} spectrum and repeat the optimal subtraction analysis with the two best-matched templates (HD 175545 and HD 165438) for each of them. For every bootstrapped copy we find the value of $v\sin i$ that results in the minimum $\chi^2$. We fit a Gaussian to the resulting distributions of $\chi^2$ values and adopt the mean and standard deviation as the best estimate of $v \sin i$ and its uncertainty.  %(see Figure \ref{fig:vsini}). 
For both templates we find that $v\sin i$ = \vsini, and the donor star contributes $\sim 45-50\%$ of the light in the NIR. 

\section{Discussion and conclusions}\label{par:conc}

For the first time we have detected absorption lines from the donor star in \gx. We observed the source 16 times between May and September 2016 with VLT/X-shooter, allowing us to reliably measure its RV curve (see Figure \ref{fig:rvcurve}). We find a systemic RV of \gam, in agreement with the value found by \citet{hynes03}, and our measurement of  $P_\mathrm{orb} =$ \porb{} confirms and refines the preferred orbital period of \citet{hynes03}. We measure the radial velocity semi-amplitude to be $K_2$ = \k2. 

\subsection{Discrepancy with Bowen blend measurement}
Our measurement of $K_2$ is significantly lower than the lower limit of $317$ \vel{} reported by \citet{hynes03} based on the Bowen blend emission lines.
One explanation for this discrepancy could be that the absorption lines in our NIR spectra are predominantly formed very close to the tip of the Roche lobe, and our $K_2$ measurement simply requires a larger K-correction than the Bowen blend measurement. Following the procedure for K-correction outlined by \citet{munoz-darias08}, taking the maximum acceptable value for $f$ with $q \approx 0.18$ based on our $v \sin i$ measurement (see Section \ref{sec:mfq}), we find a maximum correction factor of $\sim 1.7$. This could in principle reconcile our value for $K_2$ with the one found by \citet{hynes03}. This scenario would imply significant X-ray irradiation of the donor star during quiescence and that the absorption lines are stronger when we view the irradiated side of the donor star. Our spectra cover most orbital phases; we find no such variation in the presence or strength of the absorption lines. In contrast, the Bowen blend emission lines used by \citet{hynes03} are only present in part of the orbit. The total NIR flux does vary between our spectra, but not as a function of phase; it is correlated to the observing conditions. This is because the size of the extraction region is limited by the close proximity of the second star in the slit rather than by the width of the profile, which causes more flux to be lost in spectra taken under worse seeing conditions. 
These results support that our measured RVs reflect the motion of the center of mass of the donor star and we do not underestimate $K_2$.

In addition, the $K_2$ implied by the FWHM of the H$\alpha$ emission line is $206 \pm 12$ \vel, consistent with the value obtained from our dynamical study. As noted by \citet{casares15}, a caveat when using this relation is that the width of the line rises steadily after an outburst and it takes several years of quiescence to `settle down'. Because of the frequent outbursts of \gx{} it is not clear if this system ever reaches that equilibrium state. For comparison, \citet{rahoui14} obtained three spectra of \gx{} during its 2010 outburst, two while it was rising in the hard state and one after it transitioned to the soft state, and report values for the FWHM of the H$\alpha$ line of $550 \pm 30$, $704 \pm 53$, and $421 \pm 44$ at those epochs, respectively. Our FWHM of $880 \pm 20$ is significantly higher, in line with expectations. The last major outburst of \gx{} took place in December 2014, 1.5 years before our observations.  In V404 Cyg, for which \citet{casares15} shows the evolution of the H$\alpha$ FWHM in the 20 years after an outburst, the width of the line approaches its equilibrium state after $\sim 500$ days (their figure 2). This suggests that our $K_2$ measurement from this relation should be close to the real value, which further strengthens the case for a lower $K_2$ than found through the Bowen blend method.

This large discrepancy between the RV semi-amplitude measured from the Bowen blend lines and our value shows that in the case of \gx, the assumption that the narrow emission lines originate in the irradiated side of the donor star is likely wrong. Rather, the sharp Bowen blend lines must predominantly be emitted in a different location in the system with a higher Keplerian velocity or be a combination of the Keplerian motion of the star and gas flow motions. This implies that using the RV curve of the narrow Bowen blend lines may not always be a good alternative to obtaining the RV curve of the donor star by the motion of photospheric absorption lines if one wants to measure $K_2$ in BH XRBs.

\subsection{Mass function and mass ratio}\label{sec:mfq}
Our $K_2$ measurement corresponds to a mass function 
\begin{equation}
f(M) = \frac{K_2^3 P_\mathrm{orb}}{2\pi G} = \frac{M_\mathrm{x}\sin^3i}{(1+q)^2} = 1.91 \pm 0.08~\mathrm{M}_\odot
\end{equation} 
where $M_\mathrm{x}$ is the mass of the compact object and $q$ is the mass ratio between the donor star and the compact object, $ M_\mathrm{donor}/M_\mathrm{x}$. This mass function is a factor $\sim 3$ lower than derived from the Bowen emission line measurements for \gx. 

The mass ratio $q$ in a Roche-lobe overflow system is related to the projected rotational velocity and $K_2$ as 
\begin{equation}
\frac{v\sin i}{K_2} = 0.462 q^{1/3} (1+q)^{2/3}
\end{equation} 
given that the rotation of the mass donor star will be tidally locked with the orbital period \citep[e.g.][]{gies86}. We find that $v \sin i =$ \vsini, which, combined with our measurement of $K_2$, implies that $q$ = \qmeas. 

\subsection{Constraints on the inclination and BH mass}
The inclination of \gx{} has been suspected to be low rather than high based on the small separation of the double-peaked emission lines \citep{wu01}. 

\begin{figure}
\includegraphics[width=0.5\textwidth]{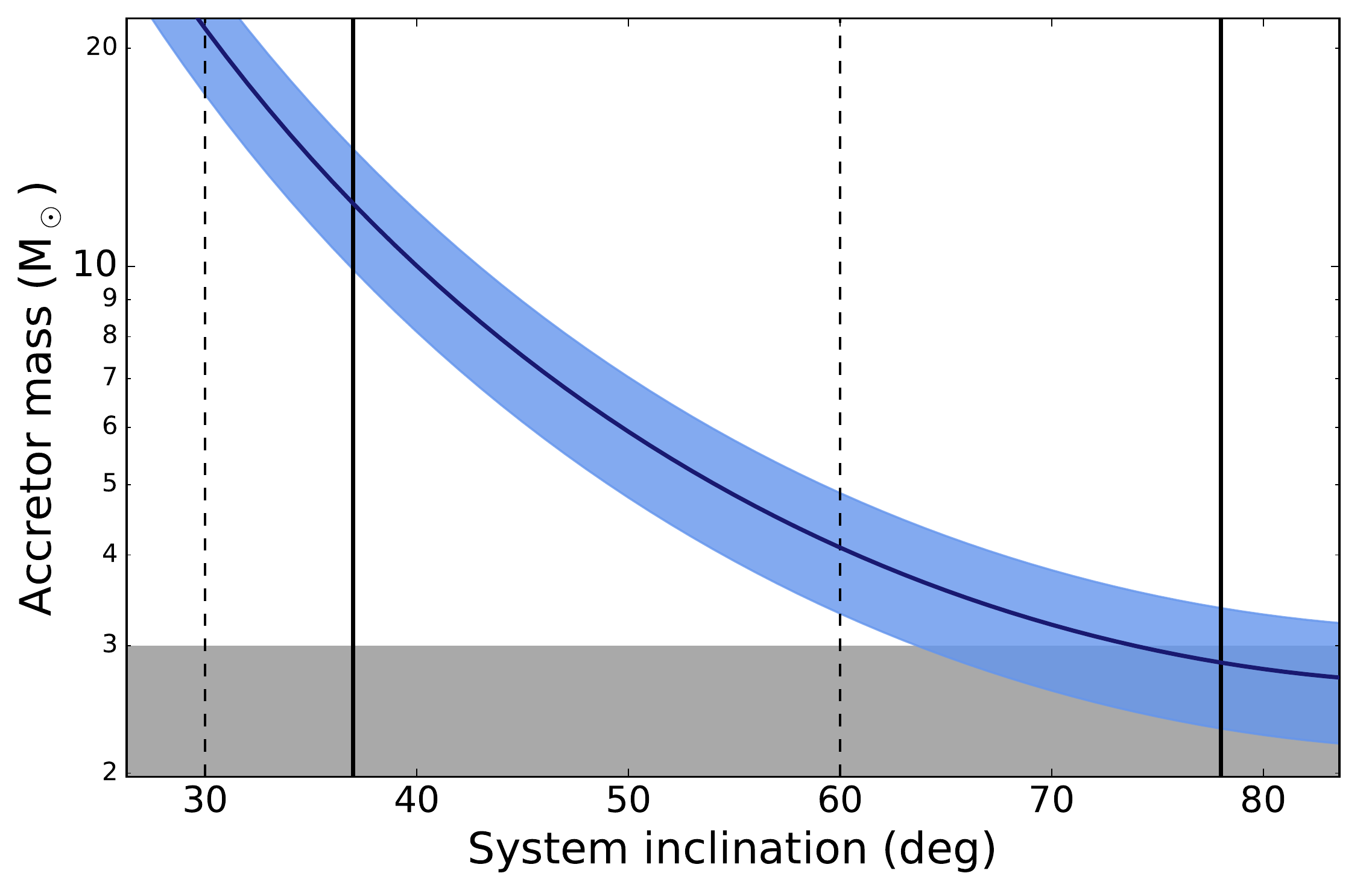}
\caption{The mass of the accretor in \gx{} as a function of the binary inclination, given that $f(M)$ = \fm{} and $q$ = \qmeas. The blue shaded area indicates the 2-$\sigma$ uncertainty. The grey shaded area indicates the theoretically allowed mass range for a neutron star \citep{lattimer12rev}. The vertical dashed lines indicate the range of inclinations found by reflection modeling, with the caveat that this is the inclination of the inner disc which does not have to be the same as the binary inclination. The solid black lines indicate the lower limit on the inclination we find from our maximum BH mass and the upper limit from the lack of eclipses.}\label{fig:massinc}
\end{figure}

Modeling of the reflection component of the X-ray spectra of \gx{} has yielded inclinations ranging from $30 - 60^\circ$ \citep[e.g.][]{fuerst15,garcia15,basak16,parker16}. An inclination in this range would lead to a BH mass between $\sim 4 - 20$ \msun (see Figure \ref{fig:massinc}). However, this method gives the inclination of the inner accretion disc which is not necessarily the same as the binary inclination if the BH spin angular momentum axis is not aligned with the binary angular momentum vector.

An upper limit to the binary inclination is provided by the fact that \gx{} does not show eclipses in its X-ray light curve; with $q$ = \qmeas, the donor star subtends an angle of $\sim 25^\circ$ as seen from the compact object \citep{paczynski71}, setting an upper limit to the inclination of $i <$ \maxinc. Combined with our $q$ value, at this inclination our mass function yields a BH mass of $2.9 \pm 0.3$ \msun, and a 95\% confidence lower limit to the mass of the compact object of $M_\mathrm{x} >$ \bhmin. 

\citet{munoz-darias08} found a maximum mass for a sub-giant donor star of 1.1 \msun. Combined with our mass ratio this implies a maximum mass for the BH of $M_\mathrm{x} = 6.1 \pm 1.7$ \msun, or a 95\% confidence upper limit of \bhmax. 
This also sets a lower limit to the binary inclination of \mininc.

A reliable measurement of the binary inclination of \gx, for example through the detection of ellipsoidal modulations, is required to determine the mass of the accretor in this system.

\subsection{Donor star properties and distance to \gx}
From our optimal subtraction analysis we conclude that the donor is most likely a K1 or K2 star, confirming the result of \citet{hynes04} and \citet{munoz-darias08}. The S/N of our spectrum is not high enough to significantly detect many of the fainter absorption lines that are present in such stars, but we clearly detect the strongest lines due to Mg I and Al I (see Figure \ref{fig:spec}). However, the strongest Si I line in the template spectra (at 1.59 $\mu$m) is not detected in our \gx{} spectrum, which may indicate that the donor has an anomalous Si abundance. Higher S/N spectra are necessary to investigate this further. The star contributes $\sim 45-50\%$ of the NIR emission. 

The distance to \gx{} is not accurately known; \citet{hynes04} found a lower limit to the distance of 6 kpc, but point out that their observations are also consistent with a much larger distance of $\sim 15$ kpc. The flux calibration of our spectra is not absolute as they were largely taken under non-photometric conditions and because the size of the extraction region is limited by the proximity of star B. However, we can derive a lower limit to the distance of \gx{} using the spectrum with the highest measured flux. This is the spectrum taken on 2016-05-22, which has an average flux density of $1.2 \times 10^{-17}$ \flxd{} in the \h{} --- we use the \h{} because it suffers the least from extinction. Taking $50\%$ as the stellar contribution in line with our optimal subtraction results, this translates into an \h{} magnitude of the donor star of $18.3$. 

For a lower limit on the distance we take the minimum mass models for the donor star from \citet{munoz-darias08}, the properties of which translate into an absolute \h{} magnitude of $\sim 2.8$. For the extinction towards \gx{} we follow \citet{munoz-darias08} and adopt $\mathrm{E}(B-V) \approx 1.2$ \citep{hynes04}, which corresponds to $A_H \approx 0.7$. Using these values we find a distance to \gx{} of $\sim 9$ kpc. The main source of uncertainty is in the flux of the donor star. 
We conservatively estimate the uncertainty on the stellar flux as $50\%$. This translates into a $25\%$ uncertainty on the distance. We thus find a lower limit on the distance of $\sim 5$ kpc, although given the fact that the X-ray activity of \gx{} favors a donor mass at the higher end of the allowed mass range \citep{munoz-darias08}, a larger distance seems more likely. These conclusions are in agreement with those of \citet{hynes04}.  
   
\subsection{Comparison with XTE J1550-564}
It is interesting to note the difference in $K_2$ between \gx{} and XTE J1550-564, which has a very similar orbital period (1.54 days) and has been suggested as a `twin' system to \gx{} \citep{munoz-darias08}. Its donor star has been identified as a K3 star \citep{orosz11}. However, XTE J1550-564 in quiescence shows an H$\alpha$ line with a FWHM of $1506 \pm 151$ \vel{} \citep{casares15}, and it has an RV semi-amplitude of $363.14 \pm 5.97$ \vel{} \citep{orosz11}. The difference between the two systems may be due to inclination: XTE J1550-564 has a relatively high inclination of $\sim 75^\circ$, and an inclination of $\sim 35^\circ$ for \gx{} would explain the difference in $K_2$. Alternatively \gx{} may contain a less massive BH than XTE J1550-564, which contains a BH with a mass in the range of 7.8 -- 15.6 \msun. The lower mass ratio in XTE J1550-564 ($q \approx 0.03$) would suggest a more massive BH if the donor star masses are similar in both systems.
  
\subsection{Conclusions}
We obtained 16 VLT/X-shooter observations of \gx{} and detected absorption lines from the donor star in the NIR spectrum, allowing us to measure the RV semi-amplitude and projected rotational velocity of the donor star for the first time. We measure $K_2$ = \k2{} and $v \sin i$ = \vsini. This implies a mass ratio $q$ = \qmeas{} and a mass function $f(M)$ = \fm. The value we find for $K_2$ is significantly lower than the value obtained using the Bowen blend method, showing that a mass function based on RV measurements of these emission lines can be incorrect. 

The donor is a K1-2 type star and we estimate that it contributes $\sim 45-50\%$ of the light in the $J$- and \h. Adopting the minimum mass model for a stripped donor star from \citet{munoz-darias08} we obtain a lower limit to the distance to \gx{} of $\sim 5$ kpc, in agreement with \citet{hynes04}.

Without the binary inclination we can only set limits to the accretor mass; we find \bhmin{} $\leq M_\mathrm{X} \leq$ \bhmax, which means that a (massive) neutron star accretor cannot be excluded. 
Although a low system inclination would still allow for a BH more massive than 5 \msun, the mass function of \gx{} is much lower than has been assumed to date and it may in fact be the first BH to fall in the `mass-gap' of $2-5$ \msun{} \citep{ozel10,farr11}.

\section*{Acknowledgments}
MH would like to thank Javier Garc{\'{\i}}a for valuable discussions. We would like to thank the ESO director for granting us DDT time for this project and the ESO staff for executing the observations. This research is based on observations made with ESO Telescopes at the La Silla Paranal Observatory under programme IDs  097.D-0915 and 297.D-5048. It also made use of data obtained from the ESO Science Archive Facility, programme ID 085.B-0751. This research made use of the software package {\sc Molly} provided by Tom Marsh. 

\bibliographystyle{aasjournal}
\bibliography{bibliography}



\end{document}